\documentstyle[12pt,twoside,fleqn,espcrc1,epsfig]{article}
\title{Direct-decay properties of charge-exchange spin giant resonances}

\author{V.A.~Rodin
$^{\rm a,c}$
and M.H.~Urin $^{\rm b,c}$\\
\ \\
$^{\rm a}$ Kernfysisch Versneller Institute, 9747AA Groningen, The Netherlands\\
$^{\rm b}$ Research Center for Nuclear Physics, Osaka
University,\\ \ \ Mihogaoka 10-1, Ibaraki, Osaka 657-0047,
Japan\\
$^{\rm c}$ Department of Theoretical Nuclear Physics,
Moscow Engineering and Physics\\ \ \ Institute, Moscow, 115409,
Russia}
\begin{document}
\maketitle

\begin{abstract}
An extended continuum-RPA approach is applied to describe direct-decay
properties of spin giant resonances in $^{208}$Bi and $^{90}$Nb.
Partial branching
ratios for direct proton decay from these resonances are evaluated.
The branching ratio for $\gamma$-decay from the spin-dipole resonance
to the Gamow-Teller resonance (main peak) is estimated.
The saturation-like
behaviour of the mean doorway-state spreading
width in $^{208}$Pb is discussed in
connection with the branching ratio for direct proton decay from
the spin-monopole
resonance and the Gamow-Teller strength distribution.
\end{abstract}

\

For the last few years, considerable efforts have been undertaken at
RCNP (Osaka) and at KVI (Groningen) to investigate direct-decay properties
of charge-exchange spin giant resonances. In particular, (i) the partial
branching ratios for proton decay from the Gamow-Teller (GT) and spin-dipole
(SD) giant resonances in $^{208}$Bi \cite{1,2} and $^{90}$Nb \cite{3} to
neutron-hole states in,
respectively, $^{207}$Pb and $^{89}$Zr have been deduced from ($^{3}$He,tp)
coincidence
experiments; (ii) the same type of experiment has been performed to search for
the
isovector spin-monopole (IVSM) and isovector monopole (IVM) giant resonances
in $^{208}$Bi \cite{4}; (iii) the branching ratio for $\gamma$-decay from
the SDR to the GTR in
the mentioned nuclei is expected to be deduced from the ($^{3}$He,t$\gamma$)
coincidence experiments \cite{5}.

These experimental studies stimulate us to develop a rather simple and
transparent approach to describe the branching ratios for
direct proton decay and $\gamma$-emission
and, as a result, to understand better the particle-hole (p-h) structure
of the mentioned spin giant resonances. The approach originally proposed in
Refs. \cite{6,7} and extended in Refs. \cite{8,9} is based on: (i) the
continuum-RPA
(CRPA) with the use of a phenomenological mean field and the Landau-Migdal p-h
interaction, as input quantities; (ii) a phenomenological description for
the coupling of (p-h)-type doorway states to many-quasiparticle configurations.
The isoscalar part of the nuclear mean field and the Landau-Migdal parameters
are ingredients of the approach. The isovector part of the nuclear mean field
and the mean Coulomb field are calculated in a self-consistent way. Some
parameters of the model used in calculations of properties of spin GRs in
$^{208}$Bi
and $^{90}$Nb (the isoscalar mean field depth $U_0$, and the Landau-Migdal
parameters $f'$ and $g'$) are listed in Table 1.
Other parameters of the isoscalar field
are taken from Ref. \cite{6}. Another ingredient of the approach is the
smearing
parameter, or the mean doorway-state spreading width, $I$. The $I$ value is
found for
each considered GR to reproduce the experimental total width of the resonance
in calculations of the energy-averaged strength function. In Table 2 the
calculated
partial branching ratios for direct
proton decay from the GTR and SDR in $^{208}$Bi \cite{8}
are given and show a rather good
agreement with the experimental data \cite{1,2}. Regarding its physical
meaning the smearing parameter is close to the imaginary part of the optical
potential. For this reason, we propose to use a saturation-like parametrization
for $I(\omega)$
($\omega$ is the excitation energy measured from the parent-nucleus
ground-state).
Such a parametrization, originally used in applying to the familiar dipole GR
\cite{9}, allows us to roughly reproduce
the $I$ values from Table 2 as well (Figure 1).

\begin{table}[htb]
\caption{
The isoscalar mean field depth $U_0$, and the Landau-Migdal
parameters $f'$ and $g'$. The calculated
relative strength $y$ of the GT main peak and the calculated branching ratio
for $\gamma$-decay from the
SDR$^{(-)}$ to the GTR (main peak) in $^{90}$Nb and $^{208}$Bi are also given.}
\begin{center}
\ \\ \
\begin{tabular}{cccccc}
\hline
Nucleus & $U_0,$ MeV& $f'$&$g'$&$y$& $b_{\gamma}\: (\times 10^{-4})$ \\
\hline
$^{90}$Nb &53.3&0.96& 0.70&0.79& 4.7 \\
$^{208}$Bi &54.1&1.0& 0.78&0.69& 2.4 \\
\hline
\end{tabular}
\end{center}
\end{table}

\begin{table}
\caption{
Calculated and experimental branching ratios for direct proton
decay from isovector spin giant resonances in $^{208}${Bi} to neutron-hole
states
in $^{207}${Pb}. All ratios are given in \%.
Contributions from decay to deep-hole states are taken
with the spectroscopic factor $S_\nu = 1$. The mean doorway-state
spreading widths and experimental total widths are also shown.}
\label{tab1}
\begin{center}
\ \\ \
\begin{tabular}{ccccccc}
\hline
\multicolumn{1}{c}{} & \multicolumn{1}{c}{} &
\multicolumn{2}{c}{GTR}
& \multicolumn{2}{c}{SDR$^{(-)}$}& \multicolumn{1}{c}{IVSMR$^{(-)}$} \\
\cline{3-7}
\multicolumn{1}{c}
{
\begin{tabular}{c}
$\nu$ \\
\phantom{gi} \\
\phantom{go}
\end{tabular}
}
& \multicolumn{1}{c}
{
\begin{tabular}{c}
$S_\nu$ \\
\phantom{go} \\
\ 
\end{tabular}
}
& \multicolumn{1}{c}
{
\begin{tabular}{c}
$b_{\nu}$ [8] \\
\end{tabular}
}
& \multicolumn{1}{c}
{
\begin{tabular}{c}
$b_{\nu}^{exp}$ [1]\\
\end{tabular}
}
& \multicolumn{1}{c}{
\begin{tabular}{c}
$b_{\nu}$ [8]\\
\end{tabular}
}
& \multicolumn{1}{c}{
\begin{tabular}{c}
$b_{\nu}^{exp}$ [2]\\
\end{tabular}
}
& \multicolumn{1}{c}
{
\begin{tabular}{c}
$b_{\nu}$ \\
\end{tabular}
}
\\
\hline
\multicolumn{1}{c}{3p$_{{\rm\frac 12}}$} & \multicolumn{1}{c}{1.0} &
\multicolumn{1}{c}{1.53} &
\multicolumn{1}{c}{1.8$\pm $0.5} 
&\multicolumn{1}{c}{0.95} &\multicolumn{1}{c}{$0.95\pm 0.28$}& 1.9 \\
\multicolumn{1}{c}{2f$_{{\rm\frac 52}}$} & \multicolumn{1}{c}{0.98} &
\multicolumn{1}{c}{1.46} & \multicolumn{1}{c}{ inc. in p$_{{\rm\frac 32}}$%
} 
& \multicolumn{1}{c}{1.94}
&\multicolumn{1}{c}{$2.10\pm 0.61$}&5.4\\
\multicolumn{1}{c}{3p$_{{\rm\frac 32}}$} & \multicolumn{1}{c}{1.0} &
\multicolumn{1}{c}{1.37} &
\multicolumn{1}{c}{2.7$\pm $0.6} 
& \multicolumn{1}{c}{2.18} & \multicolumn{1}{c}{$2.79\pm 0.81$}& 4.0\\
\multicolumn{1}{c}{1i$_{{\rm\frac{13}2}}$} & \multicolumn{1}{c}{0.91} &
\multicolumn{1}{c}{0.03} &
\multicolumn{1}{c}{0.2$\pm $0.2} & 
\multicolumn{1}{c}{3.80} & \multicolumn{1}{c}{$3.41\pm 0.98$}&21.4\\
\multicolumn{1}{c}{2f$_{{\rm\frac 72}}$} & \multicolumn{1}{c}{0.7} &
\multicolumn{1}{c}{0.09} &
\multicolumn{1}{c}{0.4$\pm $0.2} & 
\multicolumn{1}{c}{4.02} & \multicolumn{1}{c}{$3.14\pm 0.91$}&5.7\\
\multicolumn{1}{c}{1h$_{{\rm\frac 92}}$} & \multicolumn{1}{c}{0.61} &
\multicolumn{1}{c}{0.002}
& \multicolumn{1}{c}{} & 
\multicolumn{1}{c}{1.17} & \multicolumn{1}{c}{$0.97\pm 0.27$}& 3.8\\ \hline
\multicolumn{2}{r}{$\sum\limits_\nu b_\nu$} &
\multicolumn{1}{c}{4.5} &
\multicolumn{1}{c}{4.9$\pm $1.3} & 
\multicolumn{1}{c}{14.1}& \multicolumn{1}{c}{$13.4\pm 3.9$}&43.2 \\
\multicolumn{2}{r}{$b^{tot}$} &
\multicolumn{1}{c}{4.5} &
\multicolumn{1}{c}{} &
\multicolumn{1}{c}{17.3}& \multicolumn{1}{c}{}&66\\
\multicolumn{2}{r}{$I$ $(\Gamma^{exp})$} &
\multicolumn{1}{c}{3.55} & \multicolumn{1}{c}{(3.72)} &
\multicolumn{1}{c}{4.7}& \multicolumn{1}{c}{(8.4)}&4.0\\
\hline
\end{tabular}
\end{center}
\end{table}

\begin{figure}[htb]
\centerline{\epsfig{file=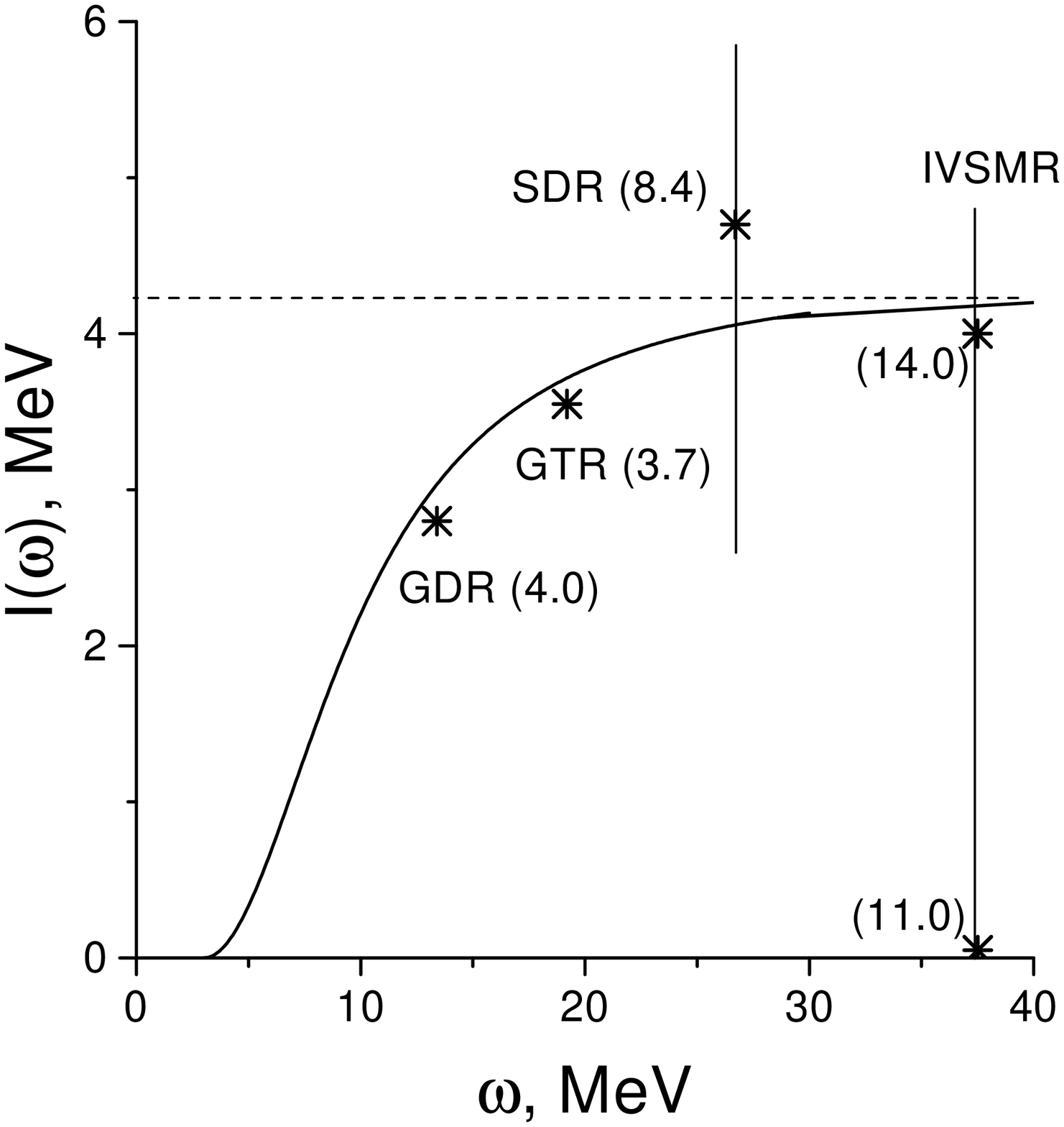,height=3.5in,width=3.5in}}
\vspace{-10pt}
\caption{The energy dependence of the smearing parameter for
$^{208}$Pb [9]. The values of doorway spreading widths
(asterisks) for spin GRs are also shown together with
the respective total widths (in parenthesis).}
\label{vrod:fig1}
\end{figure}

The reason why the IVSM/IVM giant resonances are observed in the proton
decay channel \cite{4} can be understood by comparing the total width of these
resonances in the CRPA monopole strength functions
(about 11 - 12 MeV \cite{10,11}) with
the appropriate value $I=4.2$ MeV (Figure 1). As follows from this comparison,
the proton direct-decay channel is expected to be the main decay channel
for these resonances.
To estimate the partial and total proton direct-decay
branching ratios for the IVSMR in $^{208}$Bi, we solve the respective CRPA
equations
with substitution of $\omega$ by $\omega +\frac i2 I$. In calculations with
the corrected spin-monopole one-body operator (see below) the above-mentioned
model
parameters were also used. The results in Table 2 show
that the proton direct-decay channel is the main decay channel
for the IVSMR. A similar result is also obtained for the IVMR \cite{12}.
The calculated total width
of these monopole resonances (about 14 MeV) is not in desagreement with
the experimental data \cite{4,15}. However, as follows from recent data on
searching for the IVSMR in $^{nat}$Pb and $^{90}$Zr target nuclei via the
(p,n)-reaction at 795 MeV incident energy \cite{17}, the IVSMR total width
may be larger. In this case, the larger $I$ value results in decreasing the
calculated value of $b^{tot}$ without a marked change of the relative partial
branching ratios.

We also calculated the CRPA spin-monopole strength
functions using the corrected one-body operators
$(\displaystyle\frac{r^2}{R^2}-a^2)\sigma\tau^{(-,+)}$ ($R$ is the nuclear
radius).
The parameter $a^2=0.78$ is found from the condition that the GTR (main peak)
has zero corrected
spin-monopole strength. Such a choice of the spin-monopole operator ensures
exhausting the main part of the corrected NEWSR (equals to 6.3) by the IVSMR,
which is actually
the overtone of the GTR. The calculated strength functions exhibit
resonance-like behaviour. Some parameters of the resonances in the $^{208}$Pb
(mean excitation energy $\bar\omega$, peak energy $\omega _{peak}$, FWHM, the
resonance strength x related to the respective NEWSR)
are listed in Table 3 in comparison
with the parameters from Ref. \cite{10}, where the strength functions
corresponding to the noncorrected spin-monopole one-body
operators ($a^2=0$) were calculated
using  the ``HF+Skyrme forces" mean field and the Landau-Migdal p-h interaction
with $g'= 0.78$. The main differences of the strength functions are
in exhausting the respective NEWSR and in the peak and mean energies. Note
that $\bar\omega^{exp}$ around 35.5 MeV is found in Ref. \cite{17}.

\begin{table}[htb]
\caption{Parameters of the IVSMR$^{(-,+)}$ in the $^{208}$Pb parent nucleus,
calculated in the CRPA.}
\label{tab2}
\begin{center}
\ \\ \
\begin{tabular}{ccccccccc}
\hline
\multicolumn{1}{c}{} &
\multicolumn{4}{c}{IVSMR$^{(-)}$}
& \multicolumn{4}{c}{IVSMR$^{(+)}$} \\ \cline{2-9}
\multicolumn{1}{c}
{
\begin{tabular}{c}
\end{tabular}
}
& \multicolumn{1}{c}
{
\begin{tabular}{c}
$\bar\omega$,\\ MeV \\
\end{tabular}
}
& \multicolumn{1}{c}
{
\begin{tabular}{c}
$\omega _{peak}$,\\ MeV \\
\end{tabular}
}
& \multicolumn{1}{c}
{
\begin{tabular}{c}
FWHM,\\ MeV \\
\end{tabular}
}
& \multicolumn{1}{c}
{
\begin{tabular}{c}
x \\ \ \\
\end{tabular}
}
& \multicolumn{1}{c}
{
\begin{tabular}{c}
$\bar\omega$,\\ MeV \\
\end{tabular}
}
& \multicolumn{1}{c}
{
\begin{tabular}{c}
$\omega _{peak}$,\\ MeV \\
\end{tabular}
}
& \multicolumn{1}{c}
{
\begin{tabular}{c}
FWHM,\\ MeV \\
\end{tabular}
}
& \multicolumn{1}{c}
{
\begin{tabular}{c}
x \\ \ \\
\end{tabular}
}\\
\cline{2-9}
\multicolumn{1}{c}{This work}&\multicolumn{1}{c}{
35.6}&\multicolumn{1}{c}{36.9}&
\multicolumn{1}{c}{10.0}&\multicolumn{1}{c}{0.8} & \multicolumn{1}{c}{15.1}&
\multicolumn{1}{c}{16.2}&\multicolumn{1}{c}{2.0}&\multicolumn{1}{c}{0.2} \\
\multicolumn{1}{c}{Ref. [10]} &
\multicolumn{1}{c}{44.5}&\multicolumn{1}{c}{42.5}&
\multicolumn{1}{c}{12.5}&\multicolumn{1}{c}{-}&
\multicolumn{1}{c}{17.7}&\multicolumn{1}{c}{17.8}
&\multicolumn{1}{c}{1.0}&\multicolumn{1}{c}{-}\\
\hline
\end{tabular}
\end{center}
\end{table}
Another interesting observation in our CRPA calculations is that the pygmy-
IVSMR forms the high-energy ``tail" of the GT-strength distribution in
$^{208}$Bi
and exhausts about $17 \%$ of the Ikeda sum rule (ISR) within the excitation
energy
interval $E_x=19 - 30$ MeV, while the low-energy part of the
GT-strength distribution
exhausts about $12\%$. The calculated relative strength $y$ of the GT main
peak of $69 \%$ (Table 1) is not in disagreement with the experimental
value $y^{exp}=(60\pm 15)\%$ \cite{1}.
The smeared distribution can be evaluated with the use of $I(\omega)$
shown in Figure 1.
The comparison of the calculated GT strength distribution (Figure 2) with that
deduced from the (pn) data \cite{13} allows one to verify the p-h structure of
the GT-strength distribution in $^{208}$Bi.

\begin{figure}[htb]
\centerline{\epsfig{file=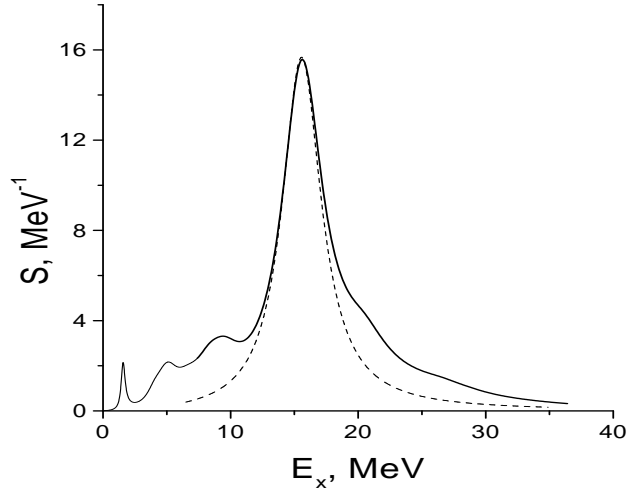,height=3.5in,width=3.5in}}
\vspace{-10pt}
\caption{The calculated GT strength distribution in $^{208}$Bi.
The total GT strength is 98\%
of the Ikeda sum rule.
The contribution of the main GT state (dotted line) is also shown.}
\label{vrod:fig2}
\end{figure}

The isospin splitting of spin GRs in nuclei having not-too-large neutron
excess is an additional element of the description. We
calculate the CRPA strength function for the GTR T$_>$-component in $^{90}$Nb
by the equation:
\begin{equation}
S_>(\omega)=\frac{1}{2T}S_{M1}(\omega -\Delta).
\end{equation}
Here $T=5$ is the isospin of the parent-nucleus ground state, $S_{M1}$ is the
CRPA
strength function corresponding to the M1 one-body operator $\sigma\tau
^{(3)}$,
and $\Delta$ is the Coulomb displacement energy. In calculations of $S_{M1}$
we put the
Landau-Migdal parameter $g=0$, because the results only slightly depend on
this parameter. The calculated mean excitation energy of the T$_>$ GTR exceeds
the
energy of the main GTR peak by 4.9 MeV. This value agrees with the
experimental GTR isospin-splitting energy 4.7 MeV \cite{14} and 4.4 MeV
\cite{3}.

We evaluate the partial branching ratios for direct proton decay from
the GTR (main peak) and SDR in $^{90}$Nb in the same way as it was done for
the spin GRs in $^{208}$Bi \cite{8}.
The results are listed in Table 4. The total branching ratio
for direct proton decay from the GTR in $^{90}$Nb is found to be
much smaller than that for the GTR in $^{208}$Bi (Table 2).
The difference is due to
the relatively low GTR excitation energy in $^{90}$Nb as compared with the
escape-proton threshold.

\begin{table}
\caption{
The calculated branching ratios for direct proton
decay from the GTR and SDR$^{(-)}$ in $^{90}${Nb} to neutron-hole
states in $^{89}${Zr} (assuming $S_\nu = 1$).
All ratios are given in \%.
The experimental total widths of the resonances is taken from Ref. [15].
}
\label{tab3}
\begin{center}
\ \\ \
\begin{tabular}{ccc}
\hline

\multicolumn{1}{c}{} & 
\multicolumn{1}{c}{GTR}
& \multicolumn{1}{c}{SDR$^{(-)}$} \\ 
\multicolumn{1}{c}
{
\begin{tabular}{c}
$\nu$ \\
\end{tabular}
}
& \multicolumn{1}{c}
{
\begin{tabular}{c}
$b_{\nu}$ \\
\end{tabular}
}
& \multicolumn{1}{c}{
\begin{tabular}{c}
$b_{\nu}$ \\
\end{tabular}
}
\\
\hline
\multicolumn{1}{c}{1g$_{\frac92}$} & 
\multicolumn{1}{c}{0.05}
&\multicolumn{1}{c}{10.0}
\\
\multicolumn{1}{c}{2p$_{\frac12}$} & 
\multicolumn{1}{c}{0.06}
& \multicolumn{1}{c}{1.6}
\\
\multicolumn{1}{c}{2p$_{\frac32}$} & 
\multicolumn{1}{c}{0.06}
& \multicolumn{1}{c}{3.6}
\\
\multicolumn{1}{c}{1f$_{\frac52}$} & 
\multicolumn{1}{c}{0.003} &
\multicolumn{1}{c}{1.7}
\\
\hline
\multicolumn{1}{r}{$\sum\limits_\nu b_\nu$} &
\multicolumn{1}{c}{0.17} &
\multicolumn{1}{c}{16.9}
\\
\multicolumn{1}{r}{$b^{tot}$} &
\multicolumn{1}{c}{0.17} &
\multicolumn{1}{c}{17.1}
\\
\multicolumn{1}{r}{$I$ $(\Gamma^{exp})$} &
\multicolumn{1}{c}{4.4 (4.4)} &
\multicolumn{1}{c}{5.0 (7.9)}
\\
\hline
\end{tabular}
\end{center}
\end{table}

Another possibility to check the p-h structure of spin GRs is to study
$\gamma$-transitions between the resonances. In connection with the expected
experimental data \cite{5} we evaluate
the branching ratios $b_\gamma$ for E1-transitions
from the SDR to the GTR (main peak) in $^{208}$Bi and $^{90}$Nb.
To simplify the
consideration, we assume that the main GT state has a wave function
close to that of the ``ideal" GT state:
\begin{equation}
|G_i\rangle =(N-Z)^{-1/2} \sum _a \sigma^{(i)} _a \tau ^{(-)}_a|0\rangle .
\end{equation}
Such an assumption allows us to express the decay probability for each SD
doorway state via its SD strength and, therefore, to reduce
the problem to the calculation
of the SD energy-averaged strength function.
The evaluated branching ratio is multiplied by the factor $y$,
which is the calculated relative (to
the ISR) strength of the main GT state. In such a way we take into account the
difference of the real main GT state from the ``ideal" one.
The calculated values of $b_\gamma$ are given in Table 1 \cite{16}.


In conclusion, we apply an extended continuum-RPA approach to evaluate: (i)
the partial branching ratios for direct proton decay from the
GTR and SDR in $^{90}$Nb, and from the IVSMR in $^{208}$Bi;
(ii) the branching ratio for $\gamma$-decay
from the SDR to the GTR (main peak) in $^{90}$Nb and $^{208}$Bi;
(iii) the GT strength distribution in $^{208}$Bi
for a wide excitation energy interval. The calculation
results are expected to be compared with the coming
experimental data.

The authors are grateful to M. Fujiwara, M.N. Harakeh and J. J\"anecke for
stimulating discussions and valuable remarks.
The authors acknowledge generous financial supports from the
``Nederlansdse organisatie voor wetenshappelijk onderzoek" (V.A.R.)
and from the RCNP COE fund (M.H.U.).


\end{document}